\documentclass[12pt,preprint]{aastex}
\usepackage{emulateapj5}

\slugcomment{Accepted by the Astrophysical Journal}

\newcommand{\Lya}{{$Ly\alpha$} }
\newcommand{\etal}{{et\thinspace al.} }
\newcommand{\cge}{{$_ >\atop{^\sim}$}}

\begin{document}
\title{ A Simple Prediction on the Surface Density of Galaxies at $z\simeq 6$}
\author{Haojing Yan\altaffilmark{1}, Rogier A. Windhorst, Stephen C. Odewahn\altaffilmark{1}, Seth H. Cohen,}
\affil{Department of Physics and Astronomy, Arizona State University,
       Tempe, AZ 85287-1504}
\email{\{Haojing.Yan, Rogier.Windhorst, Stephen.Odewahn, Seth.Cohen\}@asu.edu}
\author{Huub J. A. R\"{o}ttgering,}
\affil{Leiden Observatory, P. O. Box 9513, 2300 RA Leiden, The Netherlands}
\email{rottgeri@strw.leidenuniv.nl}
\and \author{William C. Keel}
\affil{Department of Physics and Astronomy, University of Alabama,
       Tuscaloosa, AL 35487}
\email{keel@bildad.astr.ua.edu}

\altaffiltext{1}{Visiting astronomer, Cerro Tololo Inter-American Observatory, 
National Optical Astronomy Observatory, which is operated by the Association of
Universities for Research in Astronomy, Inc. (AURA) under cooperative agreement
with the National Science Foundation.}

\begin{abstract}

Systematic surveys are being proposed to discover a
significant number of galaxies at $z\simeq 6$, which is now suggested as the
epoch when the reionization era of the Universe ends. To plan such surveys, we
need a reasonable expectation of the surface density of high redshift galaxies
at different flux limits. Here we present
a simple prediction of the surface density of $5.5 \leq z \leq 6.5$ galaxies in
the optical regime, $extrapolating$ from what is already known about galaxies 
at $z\simeq 3$.
This prediction is consistent with the results of nearly
all known searches for objects at $z\simeq 6$, giving confidence that we may
use it to plan optimal combination of survey depth and sky coverage in searching
for such objects. We suggest
that the most efficient strategy with existing ground-based facilities is
to do medium-depth ($m_{AB}\simeq 24.0$ -- $24.5$ mag), wide-field (a couple of 
square degrees) survey using a wide-field camera at a 4m-class telescope. As 
the predicted surface density at this brightness level is very sensitive to
the value of $L^{*}$, the result of such a survey can be easily used to
constrain the luminosity evolution from $z\simeq 3$ to 6.

\end{abstract}


\section{Introduction}

In the last several years, our knowledge about the Universe at high redshift
has been gradually extended to $z\simeq 6$. As of today, five galaxies at 
$z > 5.5$ (Weymann \etal 1998; Hu \etal 1999, 2002; Dawson \etal 2001) and four
quasars at $z > 5.5$ (Fan \etal 2000, 2001) have been spectroscopically 
confirmed.
The complete Gunn-Peterson trough 
detected in the $z=6.28$ SDSS quasar by Becker \etal (2001) and further 
investigation (Fan \etal 2002) led these authors to tentatively identify the
end of reionization epoch of the Universe at $z\simeq 6$.
Thus the assessment of $z\simeq 6$ galaxy number 
counts at different brightness levels will have a very direct cosmological 
impact, since it will quantify the number density of UV-emitting objects that
are the physical cause of the reionization. Several systematic surveys aimed at
discovering a significant 
number of $z\simeq 6$ galaxies have now been proposed and are being carried out.
To make such surveys efficient, we need to have a rough idea of the surface
density of galaxies at $z\simeq 6$. There are a few theoretical 
predictions on the surface density of galaxies at high redshifts, either based
on N-body simulations (cf. Weinberg \etal 1999, 2002) or based on semi-analytic 
formalisms (cf. Robinson \& Silk 2000). However, those predictions are more
qualitative than quantitative at this point, because they either are limited
by finite volume and finite resolution, or have to rely on 
several parameters which remain very uncertain in the absence of significant
amounts of actual data. Thus one may prefer not to base survey plans directly
on such predictions at the moment.

In this paper we present a simple, observational approach. We may $assume$ a 
reasonable
luminosity function for galaxies at $z\simeq 6$ by $extrapolating$ the known
results at $z\simeq 3$, which is the highest redshift where the luminosity 
function of galaxies has been quantified over a wide enough brightness range
(Steidel \etal 1999).
We will use existing data at $z$\cge 5 to constrain the normalization of this 
$extrapolated$ luminosity function. Once the $z\simeq 6$ luminosity function is
estimated in this way, the surface density can be calculated in a
straightforward manner, $i.e.$, by numerically integrating the luminosity 
function over the volume occupied by unit sky-coverage in the redshift bin of
interest. In \S 2, we present the details of such a surface density 
prediction over the range $5.5 \leq z \leq 6.5$. We compare our prediction to
all the available observations in \S 3. Comparison to two theoretical models is
made in \S 4. A summary is given in \S 5.

\section{The Predicted Surface Density of Galaxies at $5.5 \leq z \leq 6.5$}

\subsection{The Extrapolated Luminosity Function}
We start with the $z\simeq 3$ luminosity function from Steidel \etal (1999), who
merged the ground-based and the HDF-N Lyman-break galaxy samples. 
Since the $z\simeq 3$ sample used in Steidel \etal (1999) was 
selected based on the Lyman-break signature in the SED of galaxies, our 
approach makes the implicit assumption that galaxies at higher redshifts
also manifest themselves by the Lyman-break feature. Because the line-of-sight 
intervening H I absorption is more severe at higher redshifts and the effective
break moves from the 912\AA\ Lyman limit to the \Lya line at 1216\AA\ 
(cf. Madau 1995), one can expect that the Lyman-break signature is even 
stronger, therefore our assumption is likely valid. However, there might exist
a different type of galaxy whose UV-photons are completely absorbed by dust,
like the ultraluminous infrared galaxies that are known to exist at lower 
redshifts. There is no Lyman-break for such galaxies, since they have 
essentially zero flux from the UV to the optical
range. Our predictions neglect these objects, because they are also absent from 
the Lyman-break samples at $z\simeq 3$. Since we deal with the emitted UV, such
objects will not affect our conclusions.

The details of the luminosity function extrapolation from $z\simeq 3$ to $z\simeq 6$ depend on the
adopted cosmological model. Different cosmological parameters will change the
apparent magnitude of an object with a given intrinsic brightness,
and will also affect the physical volume corresponding to a given sky coverage.
We consider three cosmological models, namely, a flat universe without 
cosmological constant ($\Omega_M=1$ and $\Omega_\Lambda=0$), a flat, low mass
universe with $\Omega_M=0.3$ and $\Omega_\Lambda=0.7$, and an open universe 
without cosmological constant ($\Omega_\Lambda=0.2$ and $\Omega_\Lambda=0$). For
simplicity, we denote the three models by their $\Omega_M$ and $\Omega_\Lambda$
values as the (1, 0), (0.3, 0.7), and (0.2, 0) models, respectively. Throughout 
the paper we use a Hubble constant of $H_0=65\, km\ s^{-1}\ Mpc^{-1}$. All the
quoted magnitudes are
AB magnitudes and, unless noted otherwise, refer to the 1400 -- 1500\AA\
spectral range in the rest-frame.

The luminosity function of $z\simeq 3$ galaxies of Steidel \etal (1999; see
their section 4.3) is a Schechter function of the form
$\Phi (L) = (\Phi ^{*}/L^{*})(L/L^{*})^{\alpha}exp(-L/L^{*})$.
The $L^{*}$ galaxies at this redshift have $\mathcal{R}$-band apparent 
magnitude of $m^{*}=24.48$ mag. At $z=3.04$ (the median redshift for Steidel 
et al.'s sample), this $m^{*}$ value corresponds to $M^{*}=-20.37$, $-21.23$ 
and $-21.31$ mag in the (1, 0), (0.3, 0.7) and (0.2, 0) models, respectively. 
The $\mathcal{R}$-band has central wavelength of $\lambda_0=6930$\AA, or
$\sim 1700$\AA\ in the rest-frame at $z\simeq 3$. Since the SED of a
young non-dusty galaxy in $f_\nu$ is essentially flat from 1400 -- 1700\AA,
these magnitudes apply to the 1400 -- 1500\AA\ range as well.
As a first approximation, we also assume that there is no significant
luminosity evolution for galaxies from $z\sim 3$ to $5.5\leq z\leq 6.5$, such
that $L^{*}$ galaxies at $5.5\leq z\leq 6.5$ still have the above mentioned
absolute magnitudes. 
Possible evolution effects will be discussed in \S 5.
We noticed that Steidel \etal (1999) made a similar 
assumption in comparing the ground-based + the HDF-N sample at $z\simeq 4$ to
its counterpart at $z\simeq 3$, concluding that the observed $z\simeq 4$ galaxy
distribution was completely consistent with such an assumption at the bright 
end, and within a factor of two at the faint-end. In fact, they
suggested that the faint end mismatch was due to the genuine structure in the
small HDF-N field, rather than a true luminosity function discrepancy. We use
continuum magnitudes, ignoring the possible contribution of \Lya line emission,
for two reasons: 1) only 50\% of the galaxies in the sample of Steidel et al.
are \Lya emitters, although there was a strong bias in favor of such objects in
the spectroscopic identification, and 2) the vast majority of those \Lya 
emitters are weak in \Lya strength. We assume a slope of $\alpha =-1.6$, but
our conclusions are insensitive to the exact value.

Next we choose a normalization to fix the luminosity function, which amounts to
allowing density evolution. Observationally, this can be done by adjusting 
$\Phi ^{*}$ such that the calculated surface density, either differential or 
cumulative, matches the availible observations.
This is difficult in our case, since there is no precise observed value to be
used. There are only four published $z > 5.5$ galaxies with
spectroscopic information. These are the z=5.60
galaxy in the HDF-N (Weymann \etal 1998), the z=5.74 galaxy in the Hawaii
Survey Field SSA22 (Hu \etal 1999), and the z=5.767 
and z=5.631 galaxies in the HDF-N flanking fields (Dawson \etal 2001). 
Statistically, these results are inadequate for the purpose of normalization,
because their selection is not readily quantifiable.
Nevertheless, we can still make a rough estimate of
the cumulative surface densities of $z\geq 5.5$ galaxies in the well-studied 
HDF-N and use this value as our normalization. We consider two cases: a
cumulative surface density of 1.37 per arcmin$^2$ and 0.11 per arcmin$^2$, both
to a limit $m_{AB}=27.0$ mag. The former case (high normalization) is 
equivalent to $one$ $z\geq 5.5$ galaxy per NIC-3 field (cf. Thompson \etal 1999)
and the later one (low normalization) is equivalent to $one$ such galaxy per 
WFPC2 HDF coverage. 
Note that $m_{AB}=27.0$ 
mag is $not$ the magnitude of this z=5.60 galaxy, but an estimate of the 
selection limit for a galaxy at $z\geq 5.5$ as observed at 9100 -- 9800\AA.
This galaxy is $not$ a \Lya emitter, confirming that the magnitude in our 
prediction should be taken as referring to the continuum brightness level rather
than including this emission line.

\subsection{The Predicted Surface Densities}
Cumulative surface densities can now be calculated by numerically integrating
the luminosity function over the volume occupied by unit sky coverage in the
range
$5.5\leq z \leq 6.5$. Fig. 1 shows these results, where the cumulative surface
density is presented as the total number of galaxies per deg$^2$, whose 
brightness is brighter than a specified apparent AB magnitude. The 
predictions for the (1, 0), (0.3, 0.7) and (0.2, 0) models are plotted in solid,
long-dashed and short-dashed lines, respectively. Thick lines are used for 
the high normalization case, while thin lines are used for the low 
normalization one. The absolute magnitude scales are labeled on top of the 
figure, from bottom to the top for the (1, 0), (0.3, 0.7) and (0.2, 0) models, 
respectively. The $M^{*}$ values for these models are labeled in the legend in
the parenthesis together with the corresponding apparent magnitudes, and are
also indicated along the absolute magnitude scales at the top by upward arrows.
Numerical results are presented in Table 1, where the number
density is listed as number of galaxies per arcmin$^2$ for straightforward
use in planning future observations.

\begin{figure*}
\plotone{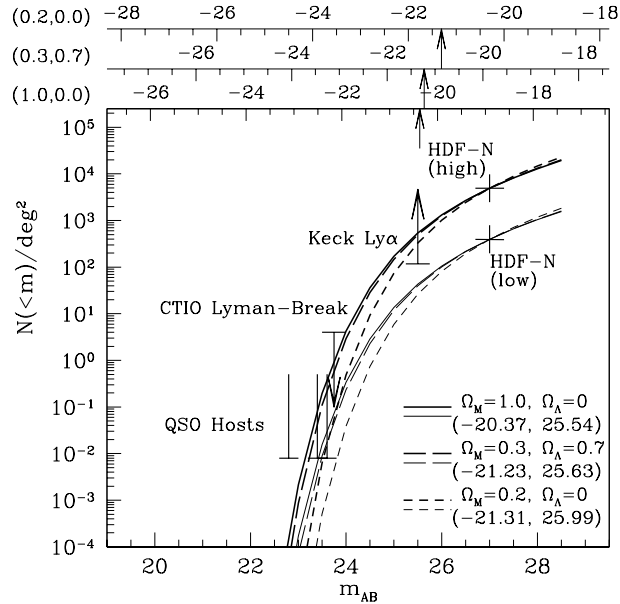}
\epsscale{0.8}
\caption{The cumulative surface density predictions of $z\simeq 6$ galaxies in
the $5.5\leq z \leq 6.5$ redshift bin, presented as the total number of 
galaxies per deg$^2$ whose brightness is brighter than a specified apparent
magnitude (defined around 9100 -- 9800\AA\, on the AB scale). The results for
the (1, 0), (0.3, 0.7) and (0.2, 0) models are 
plotted in solid lines, long-dashed lines and short-dashed lines, respectively. 
The high and low normalization cases are plotted in thick and thin lines,
respectively.
The ``$+$" labels mark the positions where the 
HDF-based normalizations are placed. The corresponding absolute magnitude 
scales are labeled on top of the figure, from bottom to the top for the (1, 0),
(0.3, 0.7) and (0.2, 0) models. The $M^{*}$ values in these three models are 
given in the legend together with the corresponding apparent magnitude scales,
and are also marked with the upward arrows on the scales. The upper limit 
derived from the preliminary data of our survey is indicated by the thick 
downward arrow, while the lower limit derived from Hu et al. (1999) is labeled
by the thick upward arrow. The indirect bright-end upper-limit inferred from the
hosts of the SDSS $5.5\leq z \leq 6.5$ quasars is indicated by ``$\bot$" from
the left to right for (1, 0), (0.3, 0.7) and (0.2, 0) models, respectively.
}
\end{figure*}

\section{Consistency Check from Limited Observations}

As a minimal consistency check, we compare our predictions to the limited
(direct and indirect) observations in hand for galaxies at $z\simeq 6$.

\subsection{The Intermediate Brightness Level\,\, {\rm ($23< m_{AB}<24$ mag)}}

Our group has started a multi-color search for $z\simeq 6$ object, aimed at 
constraining the luminosity function at this intermediate brightness level. 
Four medium-band filters, namely, $m(802nm)$, $n(848nm)$, $o(919nm)$ and
$p(974nm)$, are used in this survey. These are the four reddest of our 15 
medium-band filters which cover the entire $300$ -- $1000\, nm$ range, and are
designed to avoid the brightest and most variable night-sky lines to yield dark
background and fringe-free imaging. These four filters are ideal for 
discriminating the Lyman-break signature in the SED of galaxies at 
$5.5\leq z\leq 6.5$. Specifications of these filters can be found in the
series of papers of the BATC (Beijing-Arizona-Taiwan-Connecticut) Survey, whose 
filters are essentially the same but smaller in physical size (Fan \etal 1996;
Shang \etal 1998; Zheng \etal 1999; Yan \etal 2000). 

Our medium-band data so far allow us to put a meaningful constraint on the 
luminosity function at $z\simeq 6$. In our pilot run in June 2000, we observed
a $35'\times 35'$ area around the HDF-South in the reddest three bands
($n$, $o$ and $p$), using the MOSAIC-II array mounted at the CTIO-4m telescope.
Details of the observation will be presented elsewhere (H. Yan et al., in 
preparation). To summarize, we reached AB magnitudes of $n(848nm)$ = 23.2, 
$o(919nm)$ = 23.5 and
$p(974nm)$ = 23.0 mag at 5-$\sigma$ with $> 50\%$ completeness, and used these
images to search for drop-outs at $z\simeq 6$. We found three
candidates which were absent from the $n(848nm)$-band image but prominent in
the $o(919nm)$-band. However, since these objects are also very faint in the 
$p(974nm)$-band, imaging in bluer passbands is needed to confirm their nature.
Two out of these three objects lie within the broad band $UBVRI$ CTIO-4m BTC 
images of Palunas \etal (2000), whose survey field has about a $15'\times 15'$
overlapping region with ours. These two drop-outs are both clearly visible on
the BTC $R$-band and shortward images, but not on the I-band image. Thus these
two objects are most likely lower redshift interlopers, possibly $z=2.3$ quasars
with $MgII$ redshifted to $\sim$ 9100\AA. This meant that at best only one
positive candidate remained that could be at $z\simeq$ 5.5--6.5.

Based on these results, we can put a safe upper limit to the cumulative
number density of $5.5\leq z\leq 6.5$ objects as one object in $35'\times 35'$
down to the flux limit of $m_{AB}=23.5$ mag. This upper limit is indicated on 
Fig. 1 by an downward arrow, and is consistent with both high and low 
normalization cases in our prediction.

\subsection{The Faint End\,\, {\rm ($m_{AB} > 25$ mag)}}
A lower limit at this 
brightness level can be obtained from the $z=5.74$ galaxy of Hu \etal (1999),
assuming that its single strong emission line identification is reliable. The
galaxy's $Z$-band magnitude (free of the emission line), which is similar to
the SDSS $z'$-band magnitude, is 25.5 mag. Since their survey area is about
$390\arcsec \times 280\arcsec$, a simple lower limit estimation is $log(N)=2.07/\deg^2$ to 
$m=25.5$ mag. This lower limit is indicated on Fig. 1 as ``$\bot$", and
fits with our high normalization case but not with the low one.

Unfortunately, the two $z > 5.5$ galaxies of Dawson \etal (2001) cannot be used
as a direct constraint to the luminosity function, since no continuum
magnitude was given for either of the two. However, because their $I_{AB}$
magnitudes are all fainter than 25, their result is at least not in conflict
with our estimates above.

\subsection{Indirect Result at the Bright End\,\, {\rm ($m_{AB} \sim 23$ mag)}}
Finding any $z\simeq 6$ galaxies at these bright levels would require a 
SDSS-like all-sky survey, but extending at least one magnitude deeper, which
is a daunting, if not impossible, task with any existing facilities. However,
we can get some information in this regime by using quasar host galaxies as the
tracers of the entire galaxy population. To the extent that quasar hosts can be
taken as representatives of galaxies, it is possible to obtain a reasonable, 
although indirect, bright-end lower limit if the hosts luminosity of
the known $z\simeq 6$ quasars can be obtained. It has been long suggested that
there is a linear relation between the quasar luminosity and the $minimum$
luminosity of its host galaxy (cf.  McLeod \& Rieke 1995), which can be 
understood in the sense that a more luminous host galaxy is required to fuel a
more luminous quasar. Thus there might be a similar relation between the 
luminosity of the quasar and that of its host at higher redshift. Such a 
relation would give a luminosity function constraint from the four $z > 5.5$
quasars known, all discovered by SDSS in  $\sim$ 1500 deg$^2$ of survey area
(Fan \etal 2000, 2001). 

We construct such a relation from Bahcall \etal (1996), who imaged twenty
nearby ($z<0.3$) luminous quasars with HST, and obtained their host
luminosities. A
simple fit to their data gives $M_{host}=0.425\times M_{qso} - 11.82$ 
($rms=0.38$ mag). Using this formula, we found that the host luminosities of 
the four $z > 5.5$ SDSS quasars were almost identical. Their average absolute
magnitude at around rest-frame 
1400\AA\ is $M = -23.1$, $-23.5$ and $-23.7$ mag ($rms$ $\sim$ 0.2 mag) in the
(1, 0), (0.3, 0.7) and (0.2, 0) universe, respectively, which correspond to
apparent magnitudes of 22.8, 23.4 and 23.6 mag, respectively, at $z=6$.
The AGN unification scheme implies that if an AGN is detected, 
statistically there must be two or more galaxies which also harbor AGN activity,
which cannot be detected because obscuring matter in the surrounding torus
largely blocks our view.
In other words, since there are four $z>5.5$ quasars detected in $\sim$ 1500
deg$^2$, the total number of $z>5.5$ 
galaxies (including the four quasar hosts) is at least 12, implying a number 
density of $log(N)\geq 2.10/\deg^2$ at the above mentioned brightness levels. 
As indicated above, this number should be taken as a lower limit, because there 
could be some galaxies as luminous as the quasar hosts, but lack a massive
central black hole and thus cannot be traced by AGN. This indirect constraint
is shown by the upward arrow in Figure 1.

   There are two major uncertainties on this bright-end constraint. One is the
effect of quasar luminosity evolution. The above linear relation between the
hosts and the quasars is inferred from a local sample, yet we apply it to
$z\simeq 6$. Typical quasar luminosities have been shown to evolve with 
redshift, as $(1+z)^\beta$ with $\beta = 2.5$--4 for $z < 2.5$. This indicates
that quasars may have
once been brighter with respect to their hosts than we see today. In other
words, the hosts of the $z\simeq 6$ quasars could be fainter than we estimated
here, and therefore the limits shown in Fig. 1 could be moved more to the
right. However, since it is almost impossible to obtain a quantitative
estimate about the amplitude of such a shift, we just leave the derived
constraint as it is now. 
Another issue is whether quasar hosts are fair representatives of field 
galaxies, and therefore whether the above bright-end limit is meaningful.
Bahcall \etal (1996) speculated that the quasar hosts might not fit a 
Schechter function and might be 2.2 mag brighter than field galaxies. However,
their suggestion was based on a volume-limited sample, which is clearly not the
case for their luminous quasar sample. Furthermore, galaxy evolution clearly
cannot be neglected in comparing results from $z\approx 0.2$ to $z\approx 6$.
Net evolution of the hosts would move this constraint to higher luminosity, 
running counter to the effect of quasar luminosity evolution. 

Even in the light of these caveats, we believe that the bright-end constraint 
indicated by the 
SDSS $z>5.5$ quasars is useful. In fact, the (1, 0) model, either high or low
normalization, and the low normalization case of the (0.2, 0) model, are 
clearly not consistent with this constraint. The high normalization
case of both the (0.3, 0.7) and (0.2, 0) models, on the other hand, are
perfectly consistent with this constraint. The low normalization case of the
(0.3, 0.7) model is barely consistent with this constraint. Thus, the mere 
detection of four quasars at $z>5.5$ implies a rather high normalization to the
counts and luminosity function of galaxies at these redshifts.

\subsection{Recent Narrow-band Search}
The only reported data which seem hard to reconcile with our prediction are 
from the $z=5.7$ \Lya emitter search of Rhoads \& Malhotra (2001), who used two 
narrow-band filters to look for \Lya emission. They reported 18 robust $z$=5.7
\Lya emitters in 0.36 deg$^2$ to a 5-$\sigma$ survey limit of $m_{AB}$=24.8 mag
(average for both bands), and discussed cosmological implications of this
result. Although these objects have not yet been spectroscopically confirmed,
we compared our prediction against this result for the sake of completeness.
For simplicity, we only compare it against our high normalization (0.3, 0.7)
model.

To make such a comparison, we need both their cumulative number density
corrected for the survey volume and the 9100--9800\AA\ continuum magnitudes of
their objects.
The two narrow-band filters that they used, $815nm$ and $823nm$, are both
75\AA\ wide and can only pick up \Lya emission from $z=5.70$ -- 5.78. The 
co-moving volume of their survey is $3.27\times 10^5\, Mpc^3$ for their 0.36 
deg$^2$ sky coverage. Our model uses $5.5\leq z\leq 6.5$, giving a co-moving 
volume of $3.96\times 10^6\, Mpc^3$ for the same sky coverage. Since their 
survey volume is only 8\% of this, the 18 objects that they detected should 
only account for 8\% of the total galaxies in our prediction. In other
words, if the 18 objects were taken at their face values, the total number
of galaxies that our model would predict is 225 in our volume at $5.5\leq z\leq 6.5$. Furthermore,
the objects they found are believed to be strong \Lya emitters. Although 
spectroscopic identification is strongly biased in favor of such objects, past
studies indicate that they are only a minority of the entire population. If the 
$z\simeq 3$ Lyman-break galaxy result of Steidel \etal (1999) is universal
only 
$\sim 25\%$ of the Lyman-break galaxies have \Lya emission strong enough
to be picked up in such narrow-band searches. This means the total number of 
$z\sim 5.7$ galaxies inferred from Rhoads \& Malhotra's result is about 72, and
thus the total number of objects that our model would predict is 900. Translated
into number density, it amounts to 0.694 per arcmin$^2$. Note
that we assume their 18 objects form a complete sample down to their
detection limit, which is possibly not the case. If we made a correction for
their survey incompleteness, the inferred number density would be even higher.

The continuum magnitudes of these objects were not explicitly given in their
paper, therefore we use the equivalent width of the weakest object to
estimate their survey limit in terms of continuum magnitude.
Using the numbers from Figure 1 of Rhoads \& Malhotra (2001), we find that the
weakest emitter has $f_\nu (line) = 0.48$ $\mu Jy$ in the line, or equivalently
$m_{AB}(line)=24.70$ mag, almost approaching their detection limit. 
Using the approximation that
$f_\nu (line)/f_\nu (continuum) = (W/\Delta \lambda)+1$, where $W$ is the
equivalent width and $\Delta \lambda$ is the bandwidth, we find that 
$f_\nu (continuum) = 0.23 \mu Jy$, or $m_{AB}(continuum)=25.50$ mag. 
At this brightness level, our model gives the cumulative surface density of 
only 0.138 per arcmin$^2$. 

This comparison shows that Rhoads \& Malhotra's result gives a factor of five
times higher cumulative number density as our prediction does. Stated
differently, if their result were used as
our normalization at $m_{AB}=25.50$ mag, the number of $z\simeq 6$ objects in 
the HDF-N would be much higher ($>$ 6 times) than what have been found.
Since the Rhoads \& Malhotra's objects still need spectroscopic identification
to judge
their real nature, we shall not go further than pointing out that there
is a discrepancy.

\begin{figure*}
\figurenum{2}
\epsscale{0.8}
\plotone{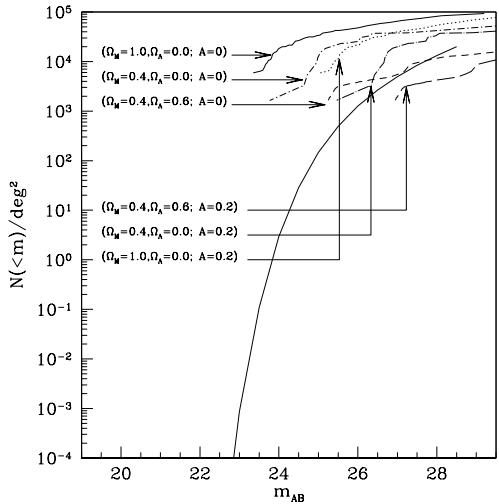}
\caption{Comparing our estimates with the N-body simulation prediction of
Weinberg et al. (2002). The high normalization case of our (0.3, 0.7) model is
plotted as the thick solid line. The thin lines of different types are the 
predictions of Weinberg et al. (2002), reproduced by reading off the numbers
from their Figure 8 and following the converting procedures given by their 
paper. Two flavors of these models are shown: one is without any dust
extinction ($A=0$ at around 1500 \AA), and the other has a dust 
extinction of $A=0.2\, mag$. Note how large the differences in the number count
predictions are with only slight differences in the assumed dust extinction
alone. Obviously only the (0.4, 0.6) model (LCDM) does not conflict with the
HDF-N result.}
\end{figure*}

\section{Compared with Theoretical Models}

We compare our estimates with the hierarchical N-body modeling results of 
Weinberg \etal (2002) in Fig. 2.
The thick solid line is our (0.3, 0.7) high normalization estimate. The thin
lines of different types are the 
predictions of Weinberg \etal (2002), reproduced by reading off the numbers
from their Figure 8 and following the converting procedures given by their 
paper. Only three of their models are reproduced here, namely, CCDM, LCDM and
OCDM, which correspond to $(\Omega_M,\Omega_\Lambda)=(1, 0)$, $(0.4, 0.6)$ and 
$(0.4, 0)$, respectively. Their choices of $\Omega_M$ and $\Omega_\Lambda$ are
slightly different from ours, but the effects of these on the
global trends are only marginal. Two flavors of these models are shown: one is 
without any dust extinction ($A=0$ at restframe 1500\AA), and the other is with
$20\%$ dust 
extinction ($A=0.2$ mag). Note that the count predictions can differ widely
solely due to different assumptions about the internal extinction.
Obviously only the (0.4, 0.6) model (LCDM) does not conflict with the HDF-N
result as summrized in our Fig. 1 and \S 2.1. Furthermore, nearly all those models predict unrealistically high surface
density at mAB$\leq 24.5$ mag, which is caused by the finite resolution in the
simulations. Trading volume for finer resolution (Weinberg \etal 1999), their
LCDM result would give a more realistic surface density at the bright end, but
again would predict too high a value at the faint end.

It is also interesting to do a comparison among the theoretical predictions.
For example, we can compare the result derived from Weinberg \etal (2002) as
shown in Figure 2 with the semi-analytic result of Robinson \& Silk (2000) as 
shown in 
their Figure 3. A direct, quantitative comparison is not meaningful, since
the result here is for $z=6$ while the highest redshift given in 
Robinson \& Silk's Figure 3 is $z=5$, and their choices of $\Omega_M$ and 
$\Omega_\Lambda$ are slightly different. But we can still compare the global 
trends, which show obvious differences.  In Robinson \& Silk's semi-analytic 
formalism, the (1, 0) model gives the lowest galaxy counts, the (0.3, 0.7) lies
in between, and the (0.2, 0) one gives the highest galaxy counts. They explained
such trends as the result caused by the dominant effects of the different
volume element $dV/dz$ and the growth factor $G$ in different cosmologies. 
A lower-density universe always has higher $dV/dz$ and larger $G$, and the 
combination of these two factors dominates the competing effect of the larger
luminosity distance which makes objects fainter in such a universe.
On the other hand, Weinberg et al.'s (1, 0) model yields the highest 
counts, (0.4, 0.6) yields the lowest counts, and (0.4, 0) is in between. 
Notice that in our approach (1, 0) gives the highest counts, (0.3, 0.7) is in
between, and (0.2, 0) gives the lowest counts. It is possible that the 
reasoning of Robinson \& Silk (2000) might only be partially true, and that
other effects which they did not consider might actually be important. For 
example, if the luminosity of $L^*$ galaxies in a high-density universe is
brighter than that can be 
produced from their current model, the number counts in different cosmologies
might have a different behavior from that shown in their paper.

\section{Discussion and Conclusion}

\subsection{The Effect of Luminosity Evolution}

The most crucial assumption made in our prediction is that the luminosity 
evolution from $z\simeq 3$ to 6 is not significant, so that the value of 
$L^{*}$ at around restframe 1400\AA\ is still the same at $z\simeq 6$ as at 
$z\simeq 3$. 
However, there are several possibilities where this condition 
could break down. There are at least two major competing effects which are
relevant. One possibility is that the merger/star-forming rate could be lower at
$z\simeq 6$, and so would bring down the value of $L^{*}$.
On the other hand, both dust extinction and metallicities could be
lower at earlier epochs as well, which would make $L^{*}$ brighter. Since these
effects tend to cancel each other out, we believe that to first order our
assumption is reasonable. 
Needless to say, the reality could be more complicated that what
we assumed.  For example, it has been suggested that $L^{*}$ evolves mildly in
the form $L^{*}\propto (1+z)^{\beta}$, where $\beta$ varies from $-1$ to $-1.5$
($e.g.$ Lanzetta \etal 1999). As described below, the surface density at the 
bright end is extremely sensitive to the $L^{*}$ value. Therefore, a direct 
comparison between this simple prediction, which serves as a first 
approximation, and any future observations can give quantitative estimates on 
how the luminosity evolves from $z\simeq 3$ to 6.
For the sake of completeness, here we
discuss how the possibilities mentioned above would affect our prediction.

The time interval between $z=3$ and $z=6$ is about 1.28 Gyr in a 
$\Omega _M=0.3$, $\Omega _\Lambda =0.7$, and $H_0=65$ universe, and is likely
only sufficient to allow one major merger ($cf.$ Makino \& Hut 1997). Hence a 
higher merger rate alone would at most make $L^{*}$ at 
$z\simeq 3$ twice as bright as at $z\simeq 6$. On the other hand, a higher 
merger rate would certainly make star formation rate higher, and this would
further contribute to the luminosity. The later effect, however, has not yet
been
well quantified. As a very rough estimation, the higher star formation rate 
could contribute another factor of few in increasing $L^{*}$ at $z\simeq 3$.
Thus the overall effect of merger/star formation rate difference would make
$L^{*}$ at $z\simeq 3$ four to six times as bright as at $z\simeq 6$, or a 
difference of 1.5--2 mag. 

The dust content and metallicities are likely to increase from $z\simeq 6$ to 3,
and these would affect $L^*$ in the opposite way. For example, one of the more
reddened
object in Steidel et al.'s sample, MS 1512-cB58, is quoted having
$E(B-V)\simeq 0.3$ mag (Pettini \etal 2001). Assuming the extinction law of
Calzetti \etal (2000), this number translates to 
$A_{1400\AA}\simeq 2.6$ mag. This means the galaxies at $z\simeq 6$ could get
brighter by 2.6 mag at most, if dust extinction is largely absent at this
redshift. The metallicity of galaxies at large redshift is again very hard to
quantify and seems to have a wide spread ($c.f.$ Nagamine \etal 2001); but in
any case, it is not likely that this effect would contribute
more than 0.5 mag increase in brightness at $z\simeq 6$ at around rest-frame
1400 \AA. 

To investigate how the surface density could be affected by the luminosity 
evolution, we also calculated the surface densities for different $M^{*}$ 
values. Specifically, we looked at the limits of the $M^{*}$ value where the
high normalization case starts to conflict with the known constraints, and where
the low normalization case begins to be consistent with those constraints.
Since $\Omega _M=0.3$ and $\Omega _\Lambda = 0.7$ are currently the most widely
accepted values, for the sake of simplicity, we will only discuss our 
(0.3, 0.7) model here. In the high normalization case, if $M^{*}$ is brighter by more
than 0.7 mag at $z\simeq 6$, the predicted counts will conflict with our 
CTIO upper limit. If $M^{*}$ is fainter by only 0.3 mag at $z\simeq 6$, on the
other hand, the 
counts will conflict with the lower limit derived from the SDSS QSO hosts by 
more than a factor of two. In the low normalization case, $M^{*}$ needs to be
brighter by at least 2.0 mag to make the counts consistent with the Keck lower
limit. In the mean time, however, such $M^{*}$ value makes the counts 
inconsistent with our CTIO upper limit, overpredicting the counts by a factor
of two. This situation further confirms that the low normalization case can
be rejected.


\subsection {Summary}

  We present a simple empirical approach to predict the galaxy surface density
at $z\simeq 6$, which $extrapolates$ the
known luminosity function of $z\simeq 3$ galaxies to $z\simeq 6$. Our approach
is based on only two observational results, namely, the observed luminosity
function of $z\simeq 3$ Lyman-break galaxies and the number of 
$5.5\leq z \leq 6.5$ galaxies in the HDF-N, and the assumption that there is no
$strong$ luminosity evolution for galaxies from $z\simeq 3$ to $z\simeq 6$.
The biggest uncertainty in our estimates comes from the normalization,
$i.e.$, the actual number density of $z\simeq 6$ galaxies in the HDF-N
down to the limit of $m_{AB}$ = 27.0 mag, for which we used one per WFPC-2
field and one per NIC-3 field as our low and
high normalization, respectively.
We checked our results against the constraints derived from all known 
observations. It seems that the low normalization case can be rejected. The
high normalization case, on the other hand, is consistent with most
constraints if the prediction is made in the (0.3, 0.7) or (1, 0) models.
The only observation with which our predictions do not agree is the narrow-band
$z=5.7$ \Lya emitter result reported by Rhoads \& Malhotra (2001), whose number
density is at least 5 times higher than our prediction. If their result were 
used as the normalization, it would suggest that the number of $z\geq 5.5$
objects in the HDF be at least 6 times as many as actually found. Since these
emitters still need future spectroscopic identification to judge their real
nature, we conclude that this is only a potential conflict. On the other hand,
a direct comparison between our prediction and any future observations
can give quantitative estimation on how the luminosity evolves at different
epochs, as indicated in the previous section.

To summarize, we believe that our
simple approach can be used to plan future surveys, where there will
always be compromise between depth and sky coverage. As our prediction
indicates, currently the most realistic way to find a significant number of 
$z\simeq 6$ galaxies with the available ground-based facilities 
is to do multi-color, medium-depth and wide-field surveys reaching continuum
$m_{AB}\sim 24.0$ -- $24.5$ mag from 8400\AA\ to the CCD Q.E. cut-off at around 1$\mu m$, and covering a couple of square degrees.
There are now several wide-field CCD cameras available at telescopes of
sufficient light-gathering power, e.g., the MOSAIC-I/II at the KPNO/CTIO 4m's,
the CFH12K at the CFHT and the Suprime-Cam at the Subaru. Carefully 
designed surveys at a 4m class telescope could possibly discover a few dozen 
$L>L^{*}$ $z\simeq 6$ galaxies within a few nights of observation (Yan \etal
2002, in preparation). On the other hand, deep, pencil-beam surveys from the 
ground are not likely to be very successful even with 8-10m class telescopes.
As Table 1 indicated, pencil-beam surveys with a few square arcmin field of 
view would have to reach at least $m_{AB}=27$ mag in the difficult spectrum
regime redder than 8400\AA\ to discover a significant number of such objects.
Since at least two bands of observation at similar depth are needed to select
drop-out candidates, the telescope time required is very costly if not
unrealistic.
In the immediate future, it should be possible to use the the Advance Camera
for Surveys, which was installed on board HST in Macrh 2002, for drop-out
searches to better constrain the luminosity function at $z\simeq 6$. 
Accessing the faint end of the luminosity function at $z\ge 6$ will be one of
the major goals that we will pursue with the Next Generation Space Telescope
(NGST) out to redshifts as high $z\simeq 9-10$, and possibly beyond.

\acknowledgements

  We are grateful to the anonymous referee for the critical comments. We thank
Drs. David Weinberg, Joseph Silk and Xiaohui Fan for helpful discussion. The
authors acknowledge support from NSF grant AST-9802963 and RAW acknowledges
support from the NGST project.  H. Yan thanks support from the Sigma Xi
Grants-in-Aid. S. Cohen would like to thank the ASU NASA Space Grant Graduate
Fellowship.

\begin{table}
\caption{Number Density Prediction for Galaxies at $5.5\leq z\leq 6.5$}
\begin{center}
\begin{tabular}{ccccccc}\tableline \tableline
      & &(1.0,\, 0.0)\tablenotemark{a}&  &(0.3,\, 0.7)\tablenotemark{a}&  &(0.2,\, 0.0)\tablenotemark{a} \\
  m   &   M    & $\Sigma$ &    M    &  $\Sigma$ &    M    & $\Sigma$ \\ \tableline
22.00 & $-23.91$ & 2.19e-14 &  $-24.86$ & 2.33e-15 & $-25.30$ & 3.02e-20 \\
22.50 & $-23.41$ & 6.83e-10 &  $-24.36$ & 1.65e-10 & $-24.80$ & 1.38e-13 \\
23.00 & $-22.91$ & 5.98e-07 &  $-23.86$ & 2.42e-07 & $-24.30$ & 2.66e-09 \\
23.50 & $-22.41$ & 5.44e-05 &  $-23.36$ & 3.06e-05 & $-23.80$ & 1.76e-06 \\
24.00 & $-21.91$ & 1.17e-03 &  $-22.86$ & 8.12e-04 & $-23.30$ & 1.34e-04 \\
24.50 & $-21.41$ & 1.00e-02 &  $-22.36$ & 7.95e-03 & $-22.80$ & 2.57e-03 \\
25.00 & $-20.91$ & 4.71e-02 &  $-21.86$ & 4.08e-02 & $-22.30$ & 2.04e-02 \\
25.50 & $-20.41$ & 1.50e-01 &  $-21.36$ & 1.38e-01 & $-21.80$ & 9.15e-02 \\
26.00 & $-19.91$ & 3.67e-01 &  $-20.86$ & 3.51e-01 & $-21.30$ & 2.82e-01 \\
26.50 & $-19.41$ & 7.53e-01 &  $-20.36$ & 7.40e-01 & $-20.80$ & 6.77e-01 \\
27.00 & $-18.91$ & 1.37e+00 &  $-19.86$ & 1.37e+00 & $-20.30$ & 1.37e+00 \\
27.50 & $-18.41$ & 2.29e+00 &  $-19.36$ & 2.32e+00 & $-19.80$ & 2.47e+00 \\
28.00 & $-17.91$ & 3.59e+00 &  $-18.86$ & 3.67e+00 & $-19.30$ & 4.09e+00 \\
28.50 & $-17.41$ & 5.40e+00 &  $-18.36$ & 5.56e+00 & $-18.80$ & 6.39e+00 \\
\tableline \\
\tableline \tableline
      & &(1.0,\, 0.0)\tablenotemark{b}& &(0.3,\, 0.7)\tablenotemark{b}& &(0.2,\, 0.0)\tablenotemark{b}\\
  m   &   M    & $\Sigma$ &    M    & $\Sigma$ &    M    & $\Sigma$ \\ \tableline
22.00 & $-$23.91 & 1.76e-15 &  $-$24.86 & 1.87e-16 &  $-$25.30 & 2.43e-21 \\     
22.50 & $-$23.41 & 5.49e-11 &  $-$24.36 & 1.32e-11 &  $-$24.80 & 1.07e-14 \\     
23.00 & $-$22.91 & 4.80e-08 &  $-$23.86 & 1.94e-08 &  $-$24.30 & 2.13e-10 \\     
23.50 & $-$22.41 & 4.36e-06 &  $-$23.36 & 2.46e-06 &  $-$23.80 & 1.41e-07 \\     
24.00 & $-$21.91 & 9.39e-05 &  $-$22.86 & 6.52e-05 &  $-$23.30 & 1.07e-05 \\     
24.50 & $-$21.41 & 8.03e-04 &  $-$22.36 & 6.38e-04 &  $-$22.80 & 2.06e-04 \\     
25.00 & $-$20.91 & 3.78e-03 &  $-$21.86 & 3.28e-03 &  $-$22.30 & 1.64e-03 \\     
25.50 & $-$20.41 & 1.20e-02 &  $-$21.36 & 1.11e-02 &  $-$21.80 & 7.35e-03 \\     
26.00 & $-$19.91 & 2.95e-02 &  $-$20.86 & 2.82e-02 &  $-$21.30 & 2.26e-02 \\     
26.50 & $-$19.41 & 6.05e-02 &  $-$20.36 & 5.94e-02 &  $-$20.80 & 5.43e-02 \\     
27.00 & $-$18.91 & 1.10e-01 &  $-$19.86 & 1.10e-01 &  $-$20.30 & 1.10e-01 \\     
27.50 & $-$18.41 & 1.84e-01 &  $-$19.36 & 1.86e-01 &  $-$19.80 & 1.98e-01 \\     
28.00 & $-$17.91 & 2.88e-01 &  $-$18.86 & 2.95e-01 &  $-$19.30 & 3.28e-01 \\     
28.50 & $-$17.41 & 4.33e-01 &  $-$18.36 & 4.46e-01 &  $-$18.80 & 5.13e-01 \\     
\tableline
\end{tabular}
\end{center}
\tablenotetext{a}{High normalization case}
\tablenotetext{b}{Low normalization case}
\tablecomments{Predictions are given for 
$(\Omega_M,\Omega_\Lambda)=(1, 0),\,(0.3, 0.7)$, and (0.2, 0) cosmologies.
$m$ is apparent magnitude and $M$ is absolute magnitude in the corresponding 
cosmology model. $\Sigma$ is cumulative number density in arcmin$^{-2}$,
calculated for the entire $5.5\leq z \leq 6.5$ bin.
}
\end{table}


\newpage
\begin{thebibliography}{}

\bibitem[Bahcall \etal (1997)]{bahcall} Bahcall, J. N., Kirhakos, S., Saxe, D. H., \& Schneider, D. P.\ 1997, \apj, 479, 642
\bibitem[]{} Calzetti, D. et al. 2000, \apj, 533, 682
\bibitem[]{} Dawson, S., Stern, D., Bunker, A. J., Spinrad, H., \& Dey, A. 2001, \aj, 122
\bibitem[]{} Fan, X., et al. 1996, \aj, 112, 628
\bibitem[]{} Fan, X., et al. 2000, \aj, 120, 1167
\bibitem[]{} Fan, X., et al. 2001, \aj, 122, 2833
\bibitem[]{} Fan, X., et al. 2002, \aj, 123, 124
\bibitem[]{} Hu, E. M., McMahon, R. G., \& Cowie, L. L. 1999, \apj, 502, L9
\bibitem[]{} Hu, E. M., et al. 2002, \apj, 568, L75
\bibitem[]{} Lanzetta, K. M., et al. 1999, in ASP Conf. Ser. 193, The Hy-Redshift Universe: Galaxy Formation and Evolution at High Redshift, ed. A. J. Bunker \& W. J. M. van Breugel (San Francisco: ASP), 544
\bibitem[]{} Madau, P. 1995, \apj, 441, 18
\bibitem[]{} McLeod, K. K. \& Rieke, G. H. 1995, \apj, 454, L77
\bibitem[]{} Makino, J. \& Hut, P. 1997, \apj, 481, 83
\bibitem[]{} Nagamine, K., Fukugita, M., Cen, R., \& Ostriker, J. 2001, \apj, 558, 497
\bibitem[]{} Palunas, P., et al. 2000, \apj, 541, 61 
\bibitem[]{} Pettini, M., et al. 2001, \apj, 554, 981
\bibitem[]{} Rhoads, J. E. \& Malhotra, S. 2001, \apjl, 563, L5
\bibitem[]{} Robinson, J. \& Silk, J. 2000, \apj, 539, 89
\bibitem[]{} Shang, Z., et al. 1998, \apj, 504, L23
\bibitem[]{} Steidel, C. C., Adelberger, K. L., Giavalisco, M., Dickinson, M., \& Pettini, M. 1999, \apj, 519, 1
\bibitem[]{} Thompson, R. I., Storrie-Lombardi, L. J., Weymann, R. J., Rieke, M. J., Schneider, G., Stobie, E., Lytle, D. 1999 \aj, 117, 17
\bibitem[]{} Weinberg, D. H., Hernquist, L., \& Katz N. 2002, \apj, 571, 15
\bibitem[]{} Weinberg, D. H. et al. 1999, in ASP Conf. Ser. 191, Photometric Redshifts and the Detection of High Redshift Galaxies, ed. R. Weymann et al. (San Francisco: ASP), 341
\bibitem[]{} Weymann, R., et al. 1998, \apj, 505, L95
\bibitem[]{} Yan, H., et al. 2000, \pasp, 112, 691
\bibitem[]{} Zheng, Z., et al. 1999, \aj, 117, 2757
\bibitem[]{} Zheng, W., et al. 2000, \aj, 120, 1607

\end{thebibliography}
\end{document}